\documentclass[aps,twocolumn,amsmath,amssymb,reprint,numbers,superscriptaddress,noeprint,longbibliography,floatfix]{revtex4-2}

\usepackage{graphicx}
\usepackage{dcolumn}
\usepackage{bm}
\usepackage[mathlines]{lineno}
\begin{document}

\title[]{Optical tuning of the diamond Fermi level measured by correlated scanning probe microscopy and quantum defect spectroscopy}

\author{Christian Pederson}
 \email{cpederso@uw.edu.}
\affiliation{University of Washington, Department of Physics, Seattle, WA, 98105, USA}

\author{Rajiv Giridharagopal}
\affiliation{University of Washington, Department of Chemistry, Seattle, WA, 98105, USA}

\author{Fang Zhao}
\affiliation{Princeton University, Department of Physics, Princeton, NJ, 08544, USA}

\author{Scott T. Dunham}
\affiliation{University of Washington, Department of Electrical and Computer Engineering, Seattle, WA, 98105, USA}

\author{Yevgeny Raitses}
\affiliation{Princeton Plasma Physics Laboratory, Princeton University, P.O. Box 451, Princeton, New Jersey 08543
}

\author{David S. Ginger}
\affiliation{University of Washington, Department of Chemistry, Seattle, WA, 98105, USA}
\affiliation{Physical Sciences Division, Pacific Northwest National Laboratory, Richland, Washington 99352, USA}

\author{Kai-Mei C. Fu}
\affiliation{University of Washington, Department of Physics, Seattle, WA, 98105, USA}
\affiliation{University of Washington, Department of Electrical and Computer Engineering, Seattle, WA, 98105, USA}
\affiliation{Physical Sciences Division, Pacific Northwest National Laboratory, Richland, Washington 99352, USA}
 
\date{\today}

\begin{abstract}
 Quantum technologies based on quantum point defects in crystals require control over the defect charge state. Here we tune the charge state of shallow nitrogen-vacancy and silicon-vacancy centers by locally oxidizing a hydrogenated surface with moderate optical excitation and simultaneous spectral monitoring. The loss of conductivity and change in work function due to oxidation are measured in atmosphere using conductive atomic force microscopy (C-AFM) and Kelvin probe force microscopy (KPFM). We correlate these scanning probe measurements with optical spectroscopy of the nitrogen-vacancy and silicon-vacancy centers created via implantation and annealing 15-25\,nm beneath the diamond surface. The observed charge state of the defects as a function of optical exposure demonstrates that laser oxidation provides a way to precisely tune the Fermi level over a range of at least 2.00\,eV. We also observe a significantly larger oxidation rate for implanted surfaces compared to unimplanted surfaces under ambient conditions. Combined with knowledge of the electron affinity of a surface, these results suggest KPFM is a powerful, high-spatial resolution technique to advance surface Fermi level engineering for charge stabilization of quantum defects.
\end{abstract}

\maketitle

\section{Introduction}

Quantum point defects in wide band gap semiconductors are utilized for quantum sensing and quantum information applications due to their long spin coherence times and spin-selective optical transitions~\cite{wolfowicz2021qgs, bassett2019qdd}. For a given application, a particular charge state of the defect is required. This charge state can be passively controlled by Fermi level engineering of the host crystal, typically through bulk impurity doping\,\cite{radishev2021inv,doi2016pnc,rose2018oei}. However, control over the Fermi level can be challenging in wide band gap semiconductors due to limits in dopability~\cite{walukiewicz1989and}. For quantum sensors and integrated quantum photonics, the defect must be close (nanometers to hundreds of nanometers) to the surface. This opens up an additional control knob, surface Fermi level engineering. Control over the surface Fermi level can be realized via active techniques such as gating~\cite{karaveli2016mnv, schreyvogel2015acs} or passive techniques such as chemical surface functionalization~\cite{hauf2011ccc,fu2010cnn,zhang2023sc,rodgers2023diamond}. Control over the surface termination is additionally attractive because it not only affects the defect charge state, it also affects defect spin~\cite{sangtawesin2019ods} and optical properties~\cite{chakravarthi2021isl}. This surface sensitivity indicates a pressing need for engineering atomically ordered surfaces with the desired electronic structure. 

A large body of existing work in diamond surface science has focused on electronic applications~\cite{crawford2021std}. Now, the impact of these surfaces on quantum point defect properties in diamond is receiving attention. Optical measurements have directly linked the ratio of fluorescence of near-surface neutral nitrogen-vacancy centers NV$^0$ and negatively charged NV$^{-}$ centers to the degree of oxidation and hydrogenation~\cite{hauf2011ccc,fu2010cnn}. Later work has shown similar switching between the neutral silicon-vacancy SiV$^0$ and negatively charged SiV$^-$ \cite{zhang2023sc}. In both cases, the surface termination was switched in a binary fashion between an oxidized surface and hydrogenated surface. However a desired defect may require an intermediate termination. For example, in the case of SiV$^0$, theoretical calculations of the formation energies suggest that near-surface SiV may not be in the desired neutral charge state for strongly hydrogenated or strongly oxidized surfaces~\cite{gali2013ais}. 

\begin{figure*}[ht]
    \centering
    \includegraphics[width=\textwidth]{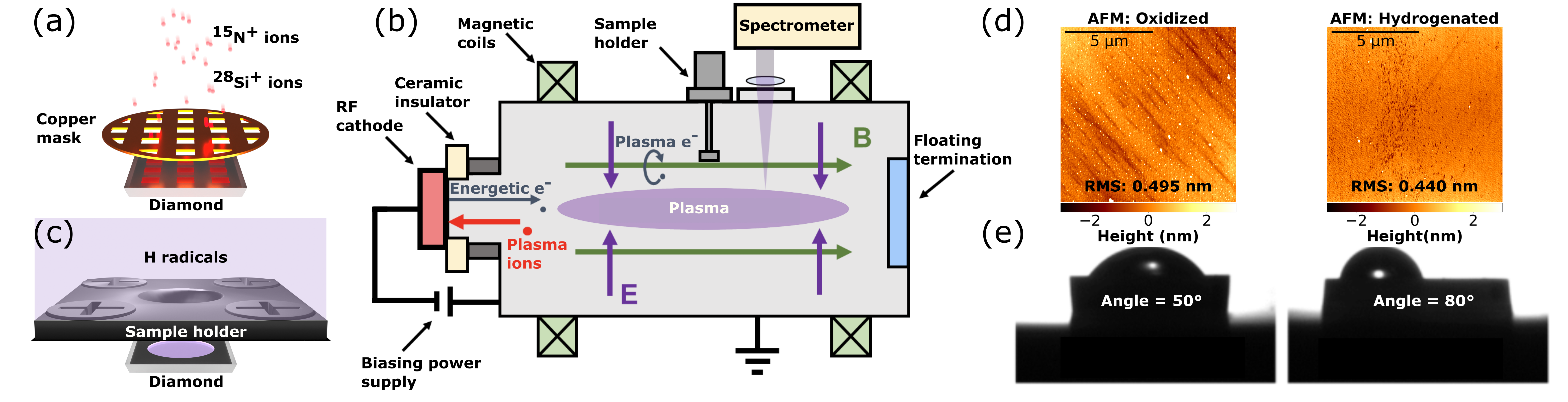}
    \caption{(a) Schematic of implantation geometry. The diamond substrate is implanted through a TEM grid resulting in squares of implanted N and Si. The squares are 28\,\textmu m (90\,\textmu m) wide in sample A (B). (b) Schematic of cold plasma reactor. (c) Schematic of the sample holder. A large circular window on the diamond surface is exposed to the hydrogen radicals, while the edges are masked. (d) AFM topography measurements before and after the hydrogenation show similar surface roughness. (e) Water wetting angle measurements before and after hydrogenation show the expected increase in hydrophobicity.
    }
    \label{fig:samples}
\end{figure*}

To date, the primary method for determining the effectiveness of the surface treatment is typically the direct optical probing of the quantum defects, in particular when the device size is small (microns)~\cite{schreyvogel2015acs} compared to standard large-area surface characterization techniques such as ultraviolet photoelectron spectroscopy (millimeters). In this work, we combine optical probing with conductive atomic force microscopy (C-AFM) and Kelvin probe force microscopy (KPFM) techniques. The scanning probe techniques work under ambient conditions, high vacuum, on fabricated electronic devices, and even under optical excitation, enabling measurements under practical experimental conditions. We find a strong correlation between the scanning probe measurements and the charge state and emission intensity of the quantum defects. We further utilize the method to study laser-assisted oxidation which provides controlled surface modification with optical spatial resolution. Finally, correlated measurements show a significant acceleration of oxidation of hydrogen-passivated surfaces in ambient conditions in the presence of implantation damage.
\section{Sample preparation and analysis techniques}
Two chemical vapor deposition diamond substrates (Element Six, ELSC grade), named A and B, were implanted with Si and N. The nominal nitrogen and boron substrate concentrations are specified as less than 5 ppb and 1 ppb, respectively. Prior to implantation, the top 5 microns of the surface were etched in an Ar/Cl RIE plasma and oxygen plasma~\cite{chakravarthi2021isl} to remove surface damage from polishing and prior implantation. We co-implant both samples with nitrogen and silicon at fluences of $10^{10}\,\text{cm}^{-2}$ and $10^{11}\,\text{cm}^{-2}$, respectively. Sample A (B) has a N$^+$ and Si$^+$ acceleration energy of 10\,keV and 19\,keV (18\,keV and 34\,keV), respectively. Implantation is performed at an angle of 7 degrees from normal through a TEM mask, resulting in implantation squares~(Fig.~\ref{fig:samples}(a)). The mean implantation depth for both impurities is 15\,nm in sample A and 25\,nm in sample B, as simulated by SRIM~\cite{ziegler2010sri}. The implantation depth is chosen such that the quantum defects lie in the band-bending region of the surface, and thus are sensitive to surface terminations~\cite{hauf2011ccc}.
The implanted N was used for forming NV centers, whose charge state is very sensitive to band-bending and relatively well understood, whereas Si was used for forming SiV in order to further our understanding of SiV charge state stabilization. Sample A (B) is annealed at 800\,$^\circ$C (1200\,$^\circ$C) for 2 hours under vacuum ($10^{-7}$\,mbar) to mobilize the vacancies produced by ion implantation and form both NV and SiV centers, and to anneal out residual implantation damage. Following the vacuum anneal, the samples are boiled in a 1:1:1 mixture of sulfuric, nitric and perchloric acid for 1 hour and subsequently annealed at ambient pressure in oxygen at 465\,$^\circ$C for 4 hours to form a starting oxidized surface~\cite{fu2010cnn}.

\begin{figure*}[ht]
    \centering
    \includegraphics[width=\textwidth]{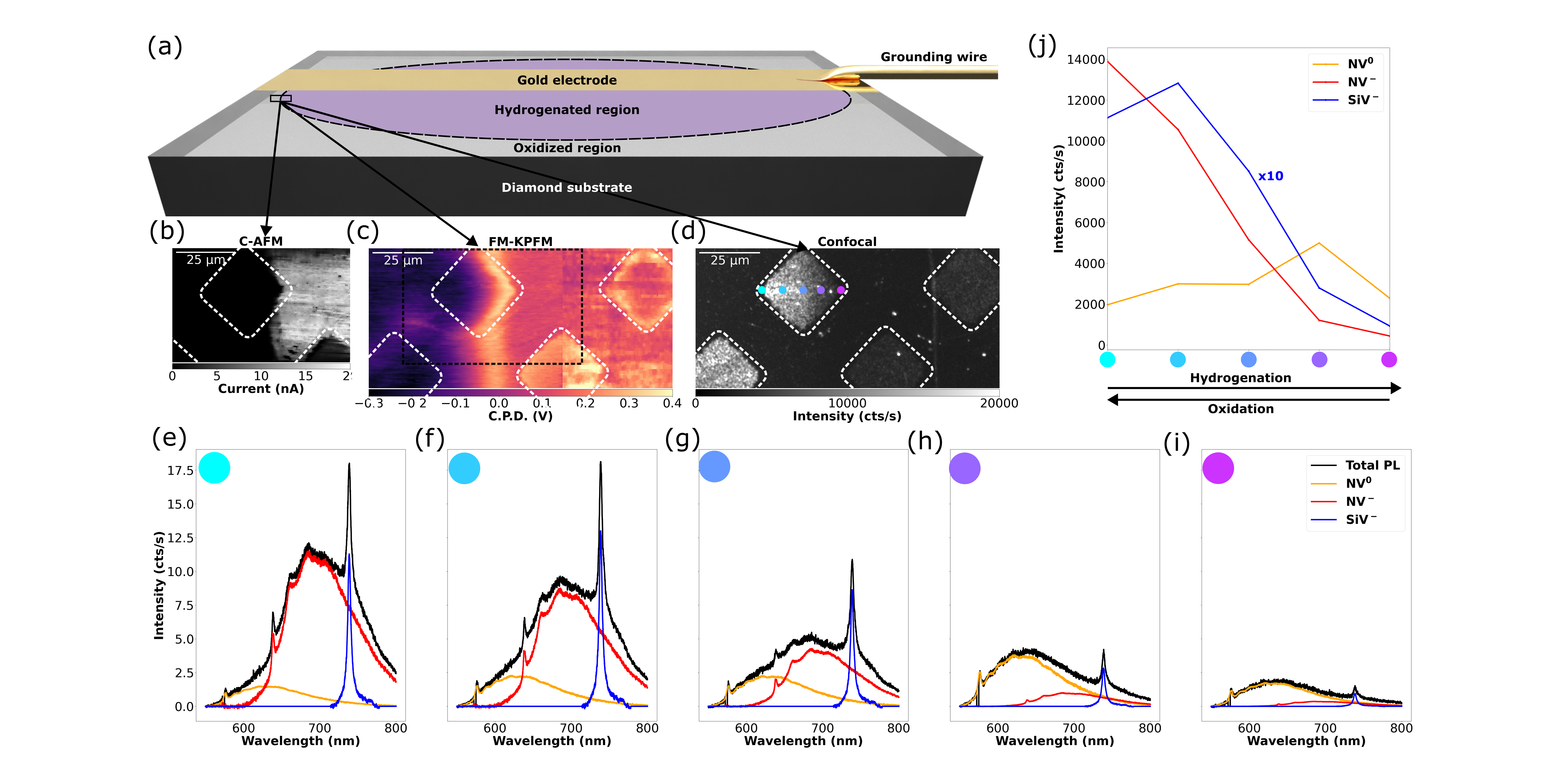}
    \caption{(a) Schematic of sample. (b) C-AFM image with a 5\,V sample bias. (c) Stitched FM-KPFM image taken with a 10\,nm lift height. A constant -70\,mV offset is applied to the right scan to match the CPD in the overlapping region. (d) Confocal photoluminescence image obtained using 1\,mW of 532\,nm excitation. (e)-(i) Photoluminescence spectra at the spots indicated in (d) obtained with 0.8\,mW of 532\,nm excitation. (j) Total PL intensity for the three defects for the spots indicated in (d).} 
    \label{fig:correlation}
\end{figure*}

Next, the sample surface was exposed to a cold hydrogen plasma treatment which was developed to achieve record hydrogen termination on graphene~\cite{Zhao2021hhc}. In conventional direct-current (DC) and radio-frequency (RF) plasma processing reactors, high energy hydrogen ions generated at the plasma-substrate interface (so-called sheath) and energetic atoms generated due to charge-exchange collisions between ions and atoms can induce substantial damage by irreversible etching and sputtering. In contrast, the cold hydrogen plasma is generated by non-thermal electrons in crossed electric and magnetic fields, so-called cross-field or ExB plasma discharge (Fig.~\ref{fig:samples}(b)). The sample is held beneath the confined plasma to expose the surface to hydrogen radicals (rather than ions). The electric field is directed inward, {\it i.e.} away from the substrate, preventing energetic ions from impinging on the substrate. AFM measurements confirm a similar RMS roughness before and after H treatment (RMS=(Fig.~\ref{fig:samples}(d)). Additionally, the sample surface becomes hydrophobic which is expected for hydrogen passivation~\cite{crawford2021std}. The sample mount exposes a circular window of the sample to the hydrogen plasma, with the sample edges masked~(Fig.~\ref{fig:samples}(c)). As we see further below, this masking allows us to probe the continuous region between the oxidized and hydrogenated surfaces. 

Confocal photoluminescence imaging and spectroscopy are performed at STP to monitor the luminescence from the NV$^0$, NV$^-$, and SiV$^-$. Excitation and oxidation are both performed with a 532\,nm laser focused to a 500\,nm diameter spot. The ratio of the NV$^-$ to NV$^0$ luminescence is used to serve as a proxy for the Fermi level in the low-excitation-power limit~\cite{hauf2011ccc}. 

For KPFM and C-AFM measurements, the ground electrode is patterned onto the face of the diamond using e-beam evaporation through a shadow mask, and wire bonded to a chip-carrier. The 140-nm-thick Au electrode is deposited on a 10-nm-thick titanium adhesion layer. This lateral geometry allows both KPFM measurements and C-AFM measurements on the same area; the use of a backing plane electrode on a thick insulating substrate would prevent C-AFM measurements. We performed KPFM using both amplitude-modulated (AM-KPFM) and frequency-modulated (FM-KPFM) methods. FM-KPFM is sensitive to the electrostatic force gradient, rather than the electrostatic force as in AM-KPFM, and therefore yields better spatial resolution and a more accurate contact potential difference (CPD), the difference between the work function of the sample and the tip. AM-KPFM is more efficient at scanning large areas quickly~\cite{melitz2011kpf}. The absolute value of the contact potential difference measured in the KPFM signal is dependent upon the material and geometry of the tip. All AFM measurements were taken with an MFP3D AFM (Oxford Instruments) using metal-coated cantilevers (KPFM: BudgetSensors ElectriTap190-G cantilevers; C-AFM: BudgetSensors CONTG-B contact-mode cantilevers). FM-KPFM was performed using custom code written in Igor Pro, as in previous work~\cite{yu2021igm}. KPFM data were taken with lift heights of typically 10 nm.

\section{Correlated PL, KPFM and C-AFM imaging}

\begin{figure*}[ht]
    \centering
    \includegraphics[width=\textwidth]{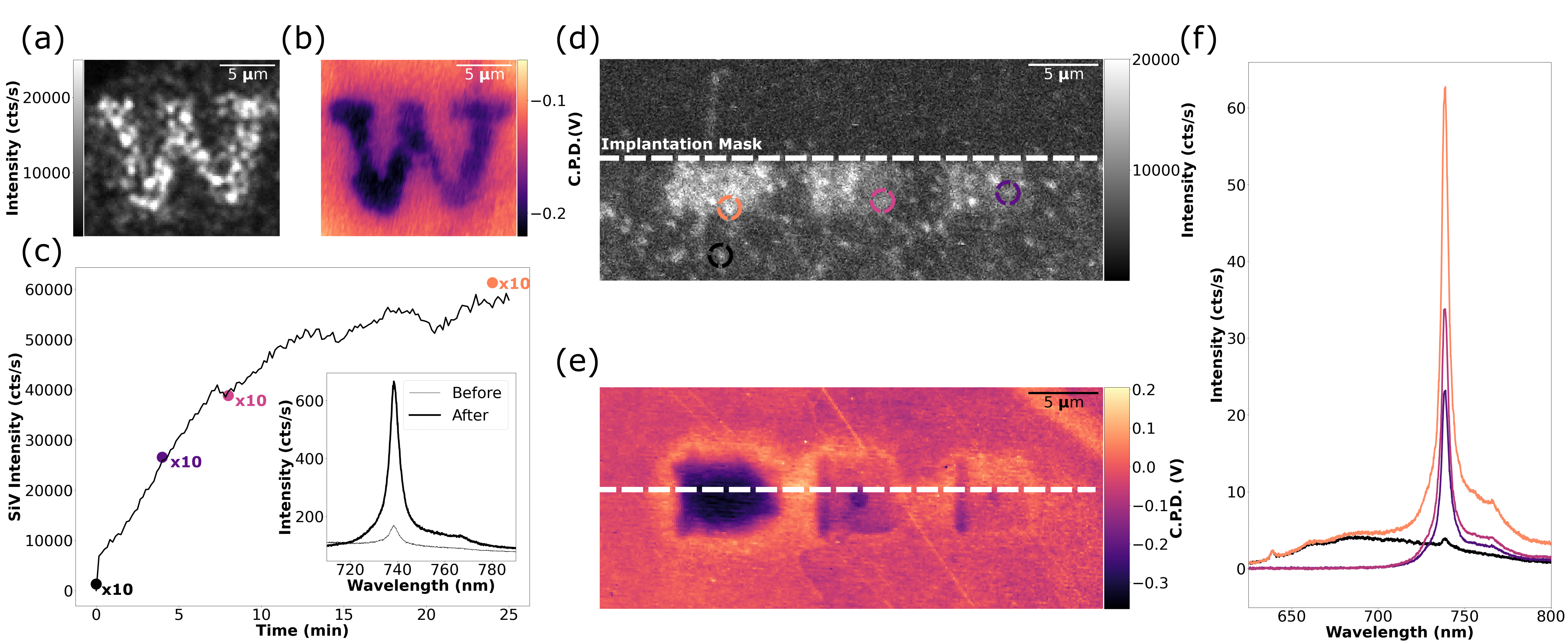}
    \caption{(a) Confocal PL image after laser-assisted oxidation. Laser-assisted oxidation is performed with 20\,mW of 532\,nm, while confocal imaging utilizes 1\,mW of 532\,nm excitation. The entire time of exposure was 10\,hours. (b) AM-KPFM image after laser-assisted oxidation. (c) Time dependence of the SiV$^-$ PL intensity using 30\,mW of excitation. Colored data points correspond to the SiV$^-$ intensity from the 3 laser-assisted exposed regions. \textit{inset:} Representative PL spectra taken at the start and end of the laser exposure used to determine total PL intensity. (d) Confocal image of three laser-assisted exposed squares using 1\,mW of 532\,nm excitation. The dashed white line denotes the implantation square boundary. PL background from deeper native NV is observed throughout the implantation square (lower half). (e) FM-KPFM image of the laser-assisted oxidized squares. (f) Photoluminescence spectra corresponding to the four marked regions in (d) and the colored data points in (c) obtained using 1\,mW of 532\,nm excitation.}
    \label{fig:laserO2}
    
\end{figure*}
We perform correlated C-AFM, KPFM and PL spectroscopy around a narrow transition region between the oxygen and hydrogen termination transition (Fig.~\ref{fig:correlation}(a)). C-AFM measurements (Fig.~\ref{fig:correlation}(b)) confirm that the hydrogen-terminated surface on the right is conductive while the left oxygen-terminated surface is insulating. This provides further verification that the hydrogen termination has pushed the Fermi level into the valence band. Additionally, a dramatic increase in resistance is observed above the implantation squares, suggesting either a loss of hydrogenation in these regions or increased scattering due to interactions between the holes and the residual implantation damage. Further measurements below support the former conclusion.
Frequency-modulated KPFM in Fig.~\ref{fig:correlation}(c) in the same region shows a 250-320\,mV potential increase from the oxidized to hydrogenated surface. The size and sign of the change in surface potential is similar to prior reports which range between 100-300\,mV~\cite{tachiki2005clm, rezek2005kfm}. Similar to C-AFM, one of the most striking features of the KPFM image is the ability to image the implantation squares. Additionally, we note that the CPD does not monotonically increase from the oxygen-terminated to hydrogen-terminated side, but rather has a maximum between the two regions. This behavior is observed even in the region that has not been implanted and will be discussed further below. 

In photoluminescence imaging, a spectral band of 653-815\,nm was collected which includes PL contributions for NV$^0$, NV$^-$ and SiV$^-$. The PL image shown in Fig.~\ref{fig:correlation}(d) shows bright PL emission under the oxidized surface, with very low fluorescence shown under the hydrogenated surface. Spatially resolved spectra across the square which lies on the termination boundary show that not only does the total intensity change (black spectra), but the relative contributions of the three defects, NV$^-$, NV$^0$ and SiV$^-$ vary as well. These relative contributions are extracted by fitting the total spectrum to a weighted sum of the individual defect spectra (Fig.~\ref{fig:correlation}(e-i)). Fig.~\ref{fig:correlation}(j) depicts the change in PL intensity for the three defects across the termination. The negatively charged defects monotonically quench with hydrogenation, while the neutral NV$^0$ PL increases but eventually quenches at the highest exposure. These results are consistent with prior reports of NV luminescence after hydrogen plasma treatment which show NV centers can be left in the neutral or even optically dark positively-charged NV$^+$ state~\cite{hauf2011ccc}, dependent on the level of hydrogenation.

\section{Optical control over the surface Fermi level}

In this section, we utilize KPFM to confirm the oxidation of the hydrogen-terminated surface via laser irradiation. Fig.~\ref{fig:laserO2}(a) depicts a confocal image of a hydrogen-terminated region of Sample A. Prior to imaging, a ``W'' was patterned on an implantation square using a 20\,mW 532\,nm laser in air focused to a 500\,nm-diameter spot. The exposed implantation square is visibly brighter, similar to the oxygen-terminated surface. AM-KPFM of the same region (Fig.~\ref{fig:laserO2}(b)) reveals that the bright PL corresponds to a decrease in the work function. In AM-KPFM, due to contributions of the work function far from the sharp tip, the magnitude of this decrease is not quantitative. These correlated measurements suggest that the level of hydrogen termination and thus the surface Fermi level can be tuned with laser irradiation. This is perhaps not too surprising given that the etching of diamond via laser-induced local oxidation has been reported at much higher laser powers~\cite{ralchenko1995fpd}. At the laser intensities used in this work, no etching is detectable via atomic-force microscopy.

We tested the hypothesis that the oxidation can be continuously tuned on sample B. In Fig.~\ref{fig:laserO2}(c), a time trace of the PL intensity from SiV$^-$ at a single excitation spot is plotted under 30\,mW excitation. The increase in intensity is observed to saturate after around 15 minutes. This relatively slow process and the optical spatial resolution enable very precise control over the Fermi level. In contrast to sample A, the spectra taken on sample B after exposure show that SiV and not NV contribute to the majority of the signal. The lower formation yield of sample B NV centers is attributed to the higher annealing temperature~\cite{orwa2011env}. 

We then exposed three 5x5 micron squares using the 30\,mW excitation, varying the exposure time to obtain 3 different photon doses. These squares overlap the edge of an implantation square, allowing us to study differences in behavior between implanted and non-implanted regions. As expected, the PL intensity only increases in the part of the square over the implanted defects (Fig.~\ref{fig:laserO2}(d)), while the entire exposed square is modified in the KPFM image (Fig.~\ref{fig:laserO2}(e)). The nonlinear response in the SiV$^-$ PL intensity with laser exposure is used to estimate the exposure from the tuning curve (Fig.~\ref{fig:laserO2}(c)). Similar to the ``W'', we observe an increase in PL intensity and a decrease in surface potential. Spectra in Fig.~\ref{fig:laserO2}(f) on the exposed squares confirm that the PL enhancement is coming from the SiV$^-$.

\section{Discussion and Outlook}

In these experiments, KPFM, correlated with photoluminescence measurements, enabled us to link the brightening of quantum defect luminescence to surface properties (rather than bulk effects). Here we discuss two additional insights. First, the work function of diamond as it transitions from hydrogenated to oxidized is non-monotonic. Second, this non-monotonic feature can be used to explain a seemingly complex spatial dependence of the work function across implantation squares. 

\begin{figure}[ht]
    \centering
    \includegraphics[width=\columnwidth]{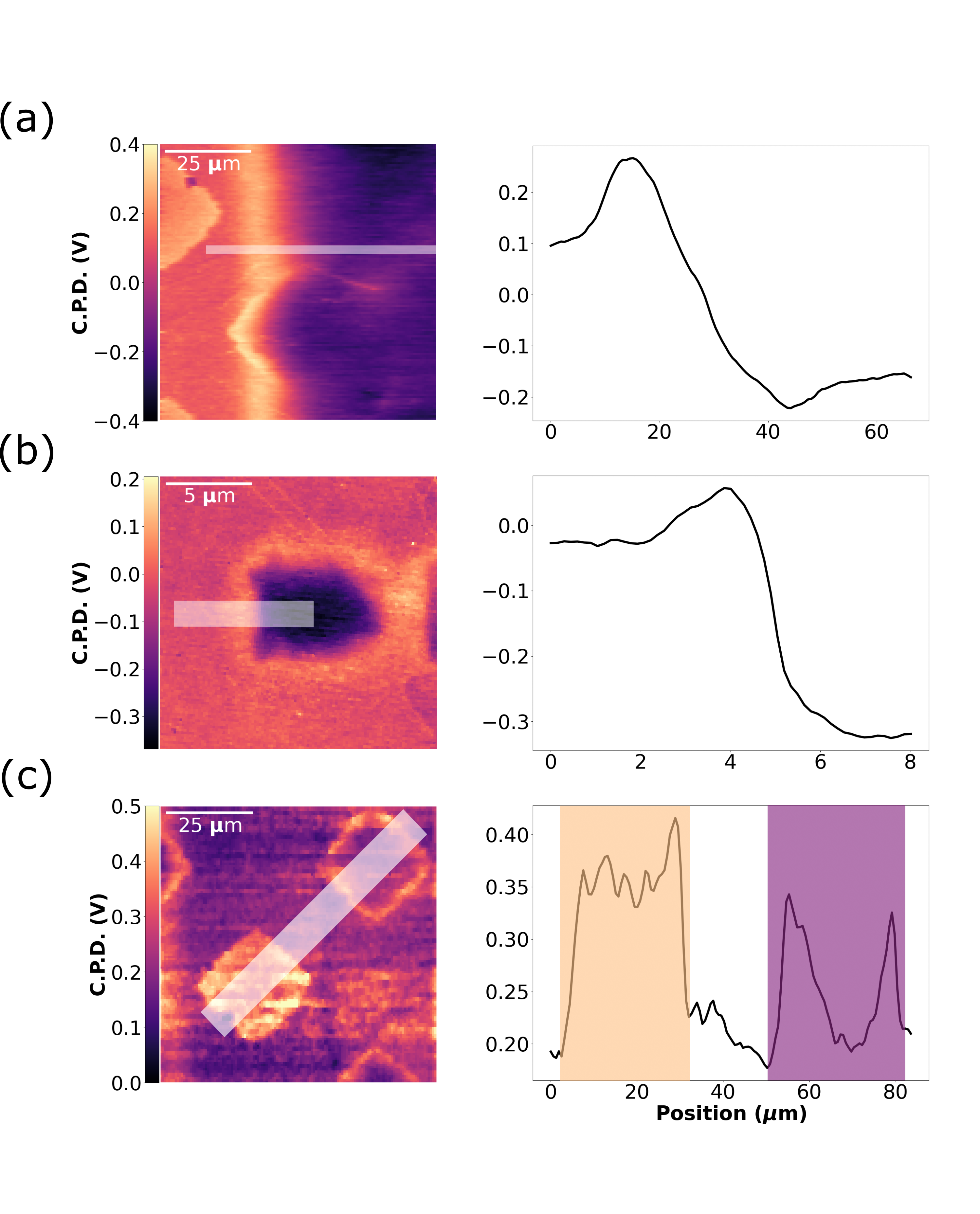}
    \caption{(a) \textit{Left:} FM-KPFM image depicting the boundary between the masked oxidized and hydrogenated diamond surface. \textit{Right:} Line scan across the unimplanted region, averaged over 2.5 microns. (b) \textit{Left:} FM-KPFM image depicting the square created by the longest exposure to the laser.\textit{Right:} Line scan across the square's boundary, averaged over 1.5 microns. (c) \textit{Left:} FM-KPFM image of two implantation squares after hydrogen passivation. \textit{Right:} Line scan across the two implantation squares, averaged over 10 microns. The implantation squares are indicated by the shaded regions.}
    \label{fig:nonmonoCPD}
\end{figure}

In Fig.~\ref{fig:nonmonoCPD}(a), a horizontal line scan of the work function is shown across the hydrogenated-oxidized boundary. For simplicity, this scan is taken in a region with no implantation. An initial increase in the work function is observed as the diamond becomes oxidized, before finally dropping 250-320\,meV below the hydrogenated level. This behavior can be understood by noting that the CPD has two contributions, the electron affinity $\chi_e$ and the Fermi level $E_F$. Specifically, the change in CPD over a material with a constant band-gap is given by $\Delta W = \Delta\chi_e-\Delta E_F$, where W is the work function, $\chi_e$ is the electron affinity, and $E_F$ is the Fermi level. The electron affinity should monotonically increase as oxygen replaces hydrogen due to the opposite polarity of the two surface dipole moments. This increase can happen spatially across the surface due to masking, or over time by our laser-induced oxidation. The Fermi level also increases as the surface is oxidized; on a hydrogenated surface, it can lie as deep as 0.7\,eV below the valence band~\cite{rezek2003flh}, while oxidized diamond is pinned around 2.0-2.4\,eV~\cite{itoh2006tmo,stacey2019eps} above the valence band. The dependence of the Fermi level on the surface termination is more complicated than the electron affinity dependence as it also depends on electron transfer between the diamond and the surface wetting layer. We hypothesize that the initial increase in the work function is caused by a Fermi level that remains pinned near the valence band despite the increasing electron affinity, while the eventual overall decrease suggests that the Fermi level eventually rises by more than the total increase in electron affinity.

The non-monotonic behavior of the CPD with oxidation results in some unique features in the KPFM images of laser-exposed and defect-implanted surfaces. In Fig.~\ref{fig:nonmonoCPD}(b) a KPFM scan of the square with the largest laser exposure is shown. At the edge of the exposed square, the increase in work function is again observed, presumably because the edge is less oxidized than inside the square. We can also explain two different characteristics of implantation squares that never received the laser oxidation treatment, illustrated in Fig.~\ref{fig:nonmonoCPD}(c). In some squares, we observe an increase in work function over the entire surface of the square, while in some squares we observe only an increase in a halo around the square. For the halo squares, the interior may exhibit a slightly higher to significantly lower work function than outside the square. These results can be explained by accelerated oxidation of the diamond surface in ambient temperatures and atmosphere due to implantation damage, even at the low implantation fluence in this study corresponding to one ion per (30\,nm)$^2$. Squares that are early in the oxidation process will have an all-positive surface, with rings forming later in the process when the interior of the square becomes more oxidized. This implantation-induced acceleration is significant; only the implanted regions in sample A show signs of oxidation after 2 years in ambient conditions. More theoretical work will be needed to elucidate the mechanism for accelerated oxidation given the importance of near-surface defects for quantum technologies.

In summary, we utilize correlated scanning-probe microscopy and photoluminescence to study surface oxidation of hydrogenated diamond surfaces. We continuously vary the hydrogenation/oxidation in two ways: by shadow masking an oxygen-terminated surface in a cold hydrogen plasma and by controlled laser-assisted oxidation. In both cases, we observe the non-monotonic increase in work function with oxidation. KPFM measurements are consistent with the observed charge state of quantum defects beneath these functionalized surfaces. Correlated measurements further uncover that the surfaces over implanted defects oxidize more quickly than pristine surfaces. Combined with knowledge of the electron affinity of a surface, these results indicate KPFM can be a powerful, high-spatial-resolution technique to advance surface Fermi level engineering for charge stabilization of quantum defects.

\section*{Acknowledgments}
The initial collaborative work for this proposal was supported by the NSF through the University of Washington Materials Research Science and Engineering Center, DMR-1719797. CP and KMF acknowledge DMR-2308979 supporting the final work. CP was partially supported by DOE Office of Science award DE-SC0022210. DSG and RG acknowledge the U.S. Department of Energy ESPM (DOE-SC0013957) for supporting the scanning probe microscopy work. Plasma treatment was supported by Princeton Collaborative Research Facility (PCRF) and funded by the U.S. Department of Energy (DOE), Office of Fusion Energy Sciences under Contract Nos. DE-AC02–09CH11466. The gold electrodes were fabricated at the Washington Nanofabrication Facility / Molecular Analysis Facility, a National Nanotechnology Coordinated Infrastructure (NNCI) site at the University of Washington with partial support from the National Science Foundation via awards NNCI-1542101 and NNCI-2025489. We thank Nathalie de Leon for helpful discussions and confirmation of the SiV charge state measurement.

\bibliography{main}

\end{document}